\documentclass[conference]{IEEEtran} 
\IEEEoverridecommandlockouts 
\usepackage{amsmath,amssymb,amsfonts}
\usepackage{algorithmic}
\usepackage{graphicx}
\usepackage{textcomp}
\usepackage{xcolor}
\usepackage{float}
\usepackage{subcaption}
\usepackage{xspace}
\usepackage{url}

\newcommand{\plantd}{PlantD\xspace}

\newcommand{\mainpoint}[1]{}
\newcommand{\honda}{Honda\xspace}
\newcommand{\cmu}{Carnegie Mellon University\xspace}
\newcommand{\modelBw}{\texttt{blocking-write}\xspace}
\newcommand{\modelNbw}{\texttt{no-blocking-write}\xspace}
\newcommand{\modelCpu}{\texttt{cpu-limited}\xspace}
\newcommand{\trafNom}{\emph{Nominal}\xspace}
\newcommand{\trafHigh}{\emph{High}\xspace}

\begin{document}

\title{PlantD: Performance, Latency ANalysis, and Testing for Data Pipelines – 
An Open Source Measurement, Testing, and Simulation Framework}

\author{
\IEEEauthorblockN{Christopher Bogart\IEEEauthorrefmark{1}, Rajeev Chhajer\IEEEauthorrefmark{2}, Baljit Singh\IEEEauthorrefmark{1},\\ Tony Fontana\IEEEauthorrefmark{2}, Majd Sakr\IEEEauthorrefmark{1}}
\IEEEauthorblockA{\IEEEauthorrefmark{1}Carnegie Mellon University, Pittsburgh, PA, USA\\
Email: \{cbogart, baljits, msakr\}@andrew.cmu.edu}
\IEEEauthorblockA{\IEEEauthorrefmark{2}Honda Research Institute USA, Inc., Columbus, OH, USA\\
Email: \{rajeev\_chhajer, tfontana\}@honda-ri.com}
}

\maketitle

\begin{abstract}
As the volume of data available from sensor-enabled devices such as vehicles expands,
it is increasingly hard for companies to make informed decisions about the cost of capturing,
processing, and storing the data from every device.  
Business teams may do detailed forecasting of costs associated with deployments and use patterns of devices that they sell, yet lack ways of forecasting the cost and performance of the data pipelines needed to support their devices. Without such forecasting,
a company's safest choice is to make worst-case capacity estimates, and pay for overprovisioned infrastructure.
Existing data pipeline benchmarking tools can measure latency, cost, and throughput as needed for development, but cannot easily close the gap in communicating the implications with business teams to inform cost forecasting. In this paper, we introduce an open-source tool, PlantD, a
harness for measuring data pipelines as they are being developed, and for interpreting that data in a business context. PlantD collects a complete suite
of metrics and visualizations, when developing or evaluating
data pipeline architectures, configurations, and business use cases. It acts as a metaphorical data pipeline wind tunnel, enabling experiments with synthetic data to characterize and compare the performance of pipelines. It then uses those results to allow modeling of expected annual cost and performance under projected real-world loads.  We describe the architecture of PlantD, walk through an example of using it to measure and compare three variants of a pipeline for processing automotive telemetry, and demonstrate how business and engineering teams can simulate scenarios together and answer "what-if" questions about the pipeline's performance under different business assumptions, allowing them to intelligently predict performance and cost measures of their critical, high-data generation infrastructure.
\end{abstract}

\begin{IEEEkeywords}
Benchmark, Performance analysis, Cloud computing, Data Pipelines
\end{IEEEkeywords}

\section{Introduction}
\label{sec:introduction}

Industry is collecting more and more data about devices.    There are currently an estimated 18 billion IoT (Internet of Things) devices in the world~\cite{colella2017}, and their number has been growing at between 20-30\% per year~\cite{dahlqvist2019growing} For example the automotive industry is predicted to collect 1-2TB of data per car by 2030 \cite{bertoncello2016monetizing}, implying that at the fleet level, a major car manufacturer may be faced with a daily data flow in the exabyte range.

As this flood of data grows, it will become increasingly unrealistic to indiscriminately collect, process, store it all forever, just in case it is needed.  Businesses will have to make hard choices, identify uses for the data and precisely what is required, with what characteristics, and at what cost to make sure the use case is worth the cost of processing and storing it.
We refer to the infrastructure built to carry all this out as a \emph{data pipeline}.

\begin{figure}
  \includegraphics[width=0.45\textwidth]{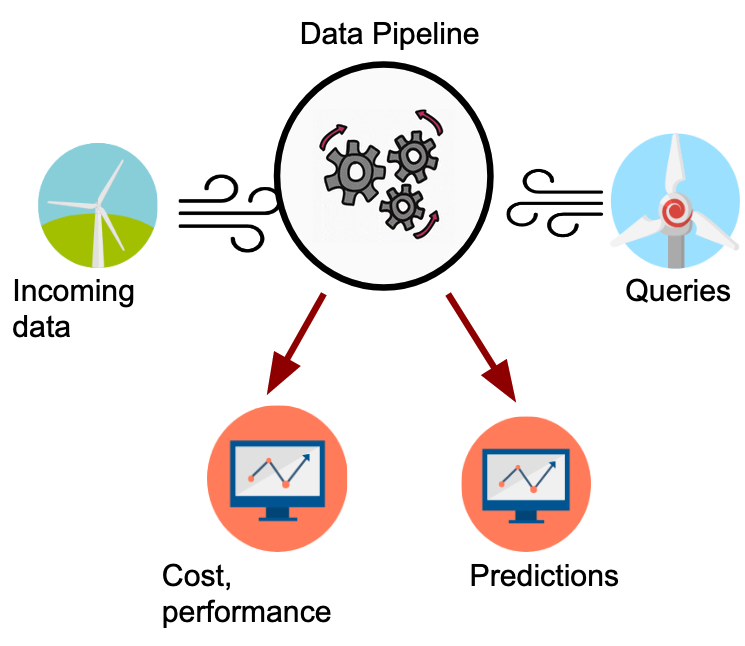}
  \caption{The data pipeline wind tunnel is a tool for instrumenting a data pipeline, subjecting it to load from incoming data and/or queries, and capturing a complete suite of metrics from it, in a form useful to engineers, managers, and business analysts.}
  \label{fig:mini-architecture}
\end{figure}

For our purposes, we consider a \emph{data pipeline} to be an infrastructural system that accepts raw data, processes it in some way, and makes it available in stored or streaming form for queries, analytics jobs, or machine learning applications. Data processing pipelines may draw from multiple data sources, produce multiple data products, and employ multiple frameworks and tools to prepare, analyze, and extract insights.  Furthermore, each use case or data product may have distinct performance requirements: for example a real time monitoring system for live road hazards will require delivering and processing telemetry with significantly less latency than, say, a system that characterizes drivers' typical driving habits to generate a monthly summary report. 

The problem of appropriate data pipeline design to meet the needs of these use cases poses two major hurdles: a business must determine the requirements of the use cases, and they must design and build a pipeline that actually meets the requirements.  

The first hurdle is to know what requirements for a data pipeline are implied by a company's proposed use cases. Business analysts need help to understand the trade-offs in scalability, accuracy, latency, and retention policies of data pipeline output, and how these choices will affect the resiliency and cost, and hence profitability, of some technology option. They need a framework for considering the relationships among a coherent set of metrics to reason about and express these needs, mapping the constraints engineers understand into a business-centric frame of reference.

The second hurdle: even if business analysts can clearly specify what business-criteria a data pipeline should meet, it is a challenge for engineers to build a pipeline to meet a set of cost and performance requirements. Each technology involved in today's complex, heterogeneous pipelines (involving components such as Kafka, Kubernetes, virtual machines, database servers, logging tools, blob storage) has unique performance characteristics depending on the volume and characteristics of the loads placed on it.  Each of these may be difficult enough to predict for a given load level and particular use case; however in combination they are in practice unknowable, since a poor estimate of the performance of one stage may interact with or inflate error in estimates of later stages.

Beyond the difficulty of planning the capacity of individual components of a pipeline is the problem of planning for a realistic system comprising many interacting components.  Data pipeline designs vary widely due to the variety of inputs, system requirements, and constraints that differ for every organization implementing a pipeline. Each unique data pipeline instance raises numerous complex questions about design, structure, feasibility, and more. With the rapidly expanding use of data pipelines, pipeline designers need tools to achieve success, and in recent years there has been an emergence of such tools as Dagger, Bugdoc, Cadet, and AdBench~\cite{Lourencco2020,Krishna2020,Rezig2020a,Bhandarkar2016}
which all attempt to ease the challenge of building a data pipeline. However, none of these tools address the full range of measurement issues pipeline designers face:
\begin{itemize}
\item Estimating cost and performance of a pipeline under a variety of assumptions, replacing the need to analyze (and trust) the cost and performance claims of a variety of interacting technologies.
\item Comparing among pipelines using very different technologies and providers, using a standard set of metrics.
\item Assembling a coherent set of tables and visualizations from the various performance data generated by the wide variety of technologies engineers might use.
\item Educating and communicating with business analysts, to help business analysts understand what data pipeline characteristics they need to consider, what various trade-offs might cost, and how to communicate business constraints back to engineers.
\end{itemize}

In this work, we introduce an open-source tool, \plantd (Performance, Latency ANalysis and Testing for Data pipelines), to address these issues. Our research team, consisting of members from \honda and \cmu, 
have designed this tool to act as a metaphorical ``data pipeline wind tunnel'', taking inspiration from the physical wind tunnel that is used by car manufacturers to study wind flow through the chassis or to evaluate the performance of various systems in the car under different conditions. In this analogy, a data pipeline is like the car in the wind tunnel, being subjected to various simulated business-specific conditions, such as number of connected devices, data formats, frequencies, and measured interactions with the external environment; to understand how they will affect the pipeline's performance, cost, and data quality.

\plantd consists of a management interface (\plantd-Studio) for tracking data schemas, pipelines, loads,  experiments, and experimental results; a load generator for sending load to a pipeline's interface(s); and a collector for gathering pipeline performance traces. It also has an business analysis engine which analyzes experimental results, builds a model of pipeline performance, and lets the user perform what-if experiments with the pipeline under different business projections of expected real-world load.

The input load generator sends synthetic load to the pipeline's input to test its ingestion and processing, but can also send queries against the pipeline's output, to test its query infrastructure. The monitoring tools pull metrics from both the load generator and the pipeline under test, to assess how well the pipeline under test manages different loads.  The data deemed most important to the pipeline designer, based on engineering or business requirements, such as on service level objectives (SLOs), then forms a benchmark for evaluating different pipeline configurations.

In the remainder of this report, we describe some of the technical challenges associated with making a general purpose data pipeline measurement tool (Sec.~\ref{sec:design_challenges}), and survey prior research into measurement and benchmarking of data pipelines (Sec.~\ref{sec:related_work}). We then detail \plantd's architecture and measurement capabilities (Sec.~\ref{sec:wind_tunnel_architecture}), in the context of a simplified implementation of a real-world pipeline in use at \honda. We then walk through a full case study to show how \plantd's analyses work; we first use it to measure a few iterations of a sample pipeline to show how \plantd enables comparison of architectural trade-offs, then we use \plantd to derive and deploy a simple mathematical model of the pipeline's performance, based on the experiment's result, to predict performance under contrasting projections of future car sales (Sec.~\ref{sec:method}).  In Results (Sec.~\ref{sec:results}) we demonstrate several what-if questions the tool allows us to answer about the suitability of the pipeline's performance under different business assumptions.  Finally, we discuss the tool's generality beyond the example, possibilities for generalizing it further, implications for research and for industrial adoption (Sec.~\ref{sec:discussion}), followed by future work and conclusion (Secs.~\ref{sec:future_work} and \ref{sec:conclusions}).

\section{Design challenges}
\label{sec:design_challenges}
There are a number of challenges associated with doing general purpose measurement of arbitrary data pipelines, involving appropriate synthesis and transmission of test load, and correct and unobtrusive measurement of performance and cost.

Synthesizing appropriate load requires generation of a wide variety of data types, file formats and packaging options.  Even synthetic data of the correct data types, ranges, and formats may still not be equivalent to real load, however, if it does not trigger the same kinds of processing in the pipeline, in the same proportions.  For example, generation of uniformly distributed latitude and longitude values will often represent points in
the ocean; a pipeline expecting to interpret this data as the location of a car would then undersample execution of code that relates the data to local law or speed limit data.

This synthesized input load then needs to be delivered to the pipeline at a well-controlled rate.  Network or storage bottlenecks in the measurement tool's load generator must be understood well enough to either scale generation appropriately, or to cap the ingestion rates that the tool promises to be able to deliver.

Measurement of the pipeline-under-test's behavior in response to this load requires some instrumentation, but the measurement tool designer must weigh the how much instrumentation they require the pipeline-under-test's engineers to insert. On one hand, highly detailed instrumentation may interfere with accurate measurement by adding network and storage burdens to the pipeline, and also may require a discouraging amount of human labor to install. On the other hand, too little instrumentation will limit how much the tool can infer about a pipeline.  For example without end-to-end tracing of packets through the pipeline, the tool cannot make accurate estimates of end-to-end latency.  Without any instrumentation at all, the tool will be unable to even tell when the pipeline is finished processing the load from an experiment.

Finally, the monetary cost of running an experiment must be captured in sufficient detail to infer the cost of running the same pipeline under expected real-world loads.  This requires teasing apart the base cost of running the infrastructure when it is idle, from the additional costs when processing various loads.  This is particularly difficult if the pipeline shares resources for which the pipeline-under-test should be allocated some share of the cost. For example if it runs on a Kubernetes cluster, it may not make sense to count cluster overhead as part of the pipeline's cost; but it may be difficult to distinguish cluster overhead by inspecting the cloud provider's billing reports.  Additionally, cloud providers often provide billing data in hourly increments, which may not align with changes in pipeline operation cost, making it hard to accurately measure the true cost of operation in a brief experiment.

\section{Related Work}
\label{sec:related_work}
\textit{The need for pipeline measurement.} Data pipelines are growing swiftly in scale.  The mobility industry, in particular, is undergoing a major disruption that will soon place great demands on cloud data processing. Four key technology trends that are driving innovation are Connectivity, Automation, Shared/Smart and Electrification (CASE) \cite{Li2021}, pointing towards a shift from individual ownership of gas-powered, hand-driven cars towards shared, electric self-driving vehicles that are constantly connected. Each of these trends have their own innovations, inventions and engineering challenges, including a massive increase of dataflow to the cloud.  Automotive telemetry data generates insights about the operation of these interrelated systems, and is providing new recurring value to customers and users. This is also a new monetization stream: a McKinsey and Company report \cite{bertoncello2016monetizing} projects that by 2030, data-enabled services will account for \$750 billion opportunity in the mobility domain, with each car easily producing as much as 1-2 TB of data per day. At the fleet level, daily data flow may extend into the exabyte range.

\textit{Definition of data pipeline.} The term \emph{data pipeline} is somewhat overloaded in the literature. For this work we use Munappy's definition of a data pipeline as ``connected chain of processes where the output of one or more processes becomes an input for another'' \cite{Munappy2020}.  Munappy's definition includes ``processes such as selection, extraction, transformation, aggregation, validation, and loading of data'', and allows for both streaming or batch pipelines, and pipelines that end either in data storage, or fed directly into machine learning or visualization engines.  
Our current work focuses on the cloud engineering aspects of pipelines; we exclude mathematical improvements such as Raman's~\cite{Raman2013} work on propagating error estimates in machine learning pipelines, or optimization of the dataflow pipelines that exist within processors. 

\textit{Generating data for testing ingestion load.}  Testing a pipeline under different conditions requires the creation of synthetic data. Soltana et al.~\cite{Soltana2017} note that data generation for pipeline testing must allow users to specify the size of the output dataset, the logical constraints that generated instances must satisfy, and the desired statistical characteristics of the generated dataset. For Soltana et al. this was achieved by using a constraint solver  and object constraint language. However, more useful to the constraints of \plantd are open source tools like SenseGen, TextBlox, and PLEDGE~\cite{Alzantot2017,Emina2012,Soltana2020}. These tools attempt to create data with statistical properties similar to the source data. 
Our proof-of-concept system starts with the simpler GoFakeIt~\cite{GoFakeIt}, to generate data with simple constraints on a specified schema.

\textit{Benchmark vs. Wind Tunnel.}  Much of the literature and tooling about data pipeline performance characteristics makes use of standard data ingestion and processing tasks, called ``benchmarks'', to characterize the performance of infrastructure technologies.  For example the ADABench~\cite{Rabl2019} benchmark is a curated collection of data pipeline tasks, datasets, processing specificaitons, and outputs; Rabl et al. used it to compare two different frameworks (Apache Spark, and scikit-learn).  The goal of such benchmarking is to compare technologies ability to implement idealized tasks.  \plantd, in contrast, is a test harness for measuring performance of a real-world data pipeline. Sfaxi and Aissa's Babel~\cite{Sfaxi2021}, is closer to \plantd in intent: it is described as a ``benchmarking platform'', and is much like our ``wind tunnel''; it measures end-to-end performance of any big-data data pipeline, although unlike \plantd it does only black-box measurements, and does not attempt to measure cost.

\textit{Monitoring data pipeline performance.} Many tools proposed or in use for \emph{monitoring} data pipelines are designed to target a specific aspect or type of data pipeline. Some tools treat the pipeline like a machine learning application, monitoring factors like prediction error or efficiency in order to apply hyperparameter tuning techniques, to optimize some particular aspect of the system~\cite{Quemy2019,Olson2016,Anderson2017}. Researchers have proposed tools such as  Bhandarkar's AdBench~\cite{Bhandarkar2016}, and Samant et al.'s automated resource scaling techniques~\cite{Samant2019}, to stress-test and monitor individual stages of a pipeline, and use this to infer or control end-to-end performance. 
Emerging technologies like BugDoc, CADET and Dagger provide automated methods to facilitate the debugging of pipelines~\cite{Lourencco2020,Krishna2020,Rezig2020a,Rezig2020b}. 

The \plantd project attempts to provide a broad platform and standard set of metrics that transcend particular technology choices and engineering problems, so that as engineers use it to make choices and solve problems, the wind tunnel also exposes the engineers' progress in a way that can be made visible and understandable for the whole organization.

\section{Using \plantd}

\begin{figure*}
  \includegraphics[width=\textwidth]{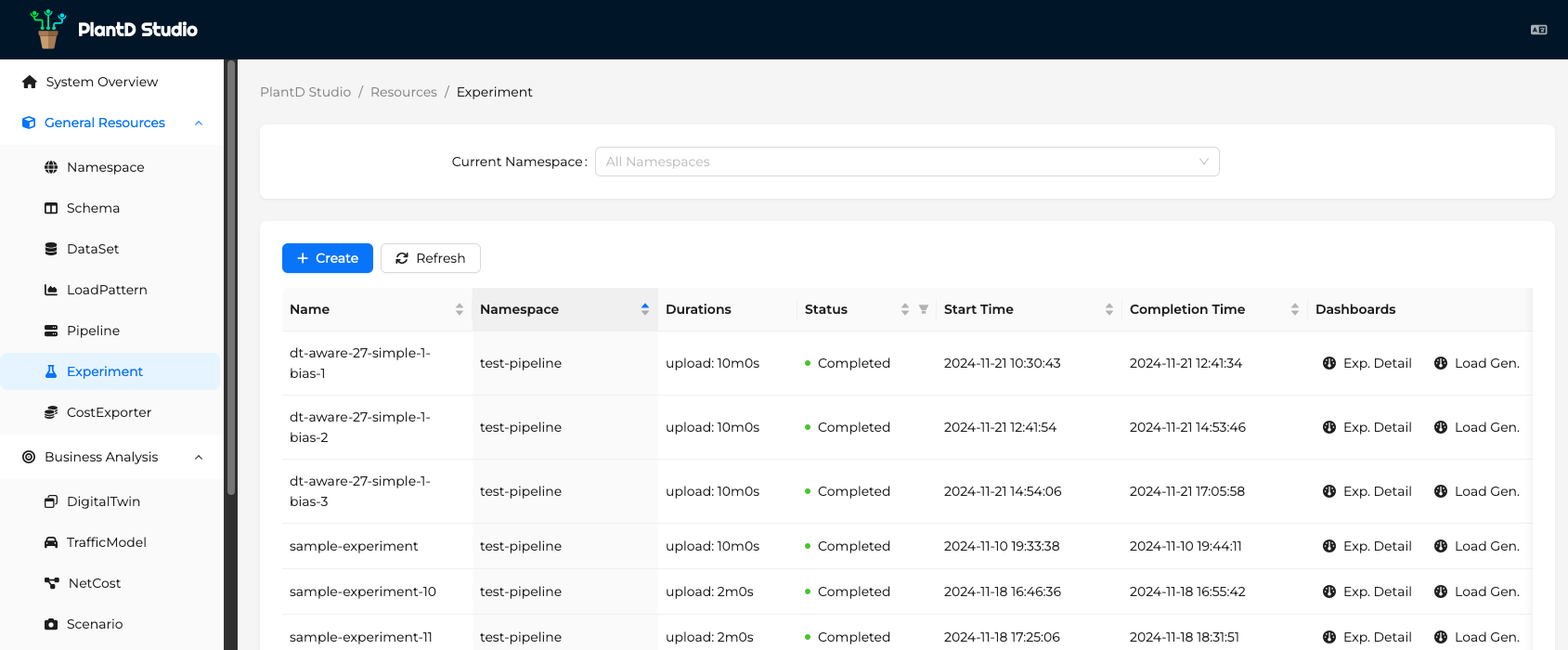}
  \caption{Screenshot of user interface, showing recently run experiments and their status.}
  \label{fig:ui}
\end{figure*}

To use \plantd (\url{https://plantd.org}) to measure and track progress of the performance of a data pipeline (we refer to this as the ``pipeline-under-test''), an engineering team must first set up a Kubernetes cluster for \plantd (or use an existing one) and install \plantd's resources on it from its GitHub repository at \url{https://github.com/CarnegieMellon-PlantD/PlantD-operator}, following instructions at \url{https://plantd.org/docs/tutorial/installation}.  They would then take the following steps:

\textbf{Prepare the pipeline for monitoring} by instrumenting the \emph{span} of each stage of the pipeline, using OpenTelemetry~\cite{OTel}. OpenTelemetry is an open source observability framework that enables collection and management of traces, metrics, and logs. Then install a \plantd-provided collector module, which will convert trace spans into metrics, and send them to the Prometheus~\cite{Prometheus} time-series database for later visualization and analysis.  \plantd's distribution comes with a toy pipeline that demonstrates how this instrumentation can be added. 

After preparing the pipeline itself, engineers can complete remaining steps in the PlantD-studio web interface.

\textbf{Create a dataset} by specifying through PlantD-studio how many of each schema to generate, and how to format and package them (e.g. zipped CSV files). Schemas are entered by listing data fields, with constraints on their values, as configuration for PlantD's random data generator.  

\textbf{Create a load pattern}: i.e. the duration and data rate of your experiment. Data rate can linearly increase, decrease, or stay steady, over segments of any length, to approximate any load curve. A useful pattern is to slowly ramp up the rate from zero up to some rate significantly above what you believe the pipeline will handle, in order to identify its nominal throughput and its behavior when that is violated.

\textbf{Describe the pipeline endpoint(s)} by the URL and protocol for data ingestion, any tags or namespace (for identifying its cost) and a metrics endpoint. The pipeline need not run on Kubernetes, or even the same cloud service as \plantd, as long as its endpoints are accessible.

\textbf{Create an experiment} by specifying a Load Pattern, Dataset, Pipeline, and pipeline endpoint, and whether the experiment should start immediately or at some scheduled time. The interface allows specification of multiple endpoints -- if an experiment makes use of these, each will need its own dataset and load pattern reference, and a pointer to the particular pipeline endpoint. 

When an experiment starts, PlantD will mark the experiment's pipeline as "engaged", until the end of the experiment.  The system will check that the pipeline is reachable, and send synthetic data matching the schema to the pipeline's endpoint at the requested rates. It will not start another experiment until the first one is done.

\textbf{View results} live, by clicking the links on the list of experiments. One link shows the load data being fed into the experiment, and another shows throughput and latency of data passing through each instrumented stage of the pipeline as they are sent. The data can be reviewed later to compare different experiments, or for posthoc analysis in the Business Analysis features of PlantD.

\textbf{Model and compare results} by selecting one or more experiments, and designating them as training data for constructing a simulation model of the pipeline. An engineer or business analyst can then use this model to simulate a year-long load pattern which might vary by season, day of week, and time of day. The simulation predicts annual cost, performance, and frequency of Service Level Objective violations.

\section{\plantd Architecture}
\label{sec:wind_tunnel_architecture}

\begin{figure*}
  \includegraphics[width=\textwidth]{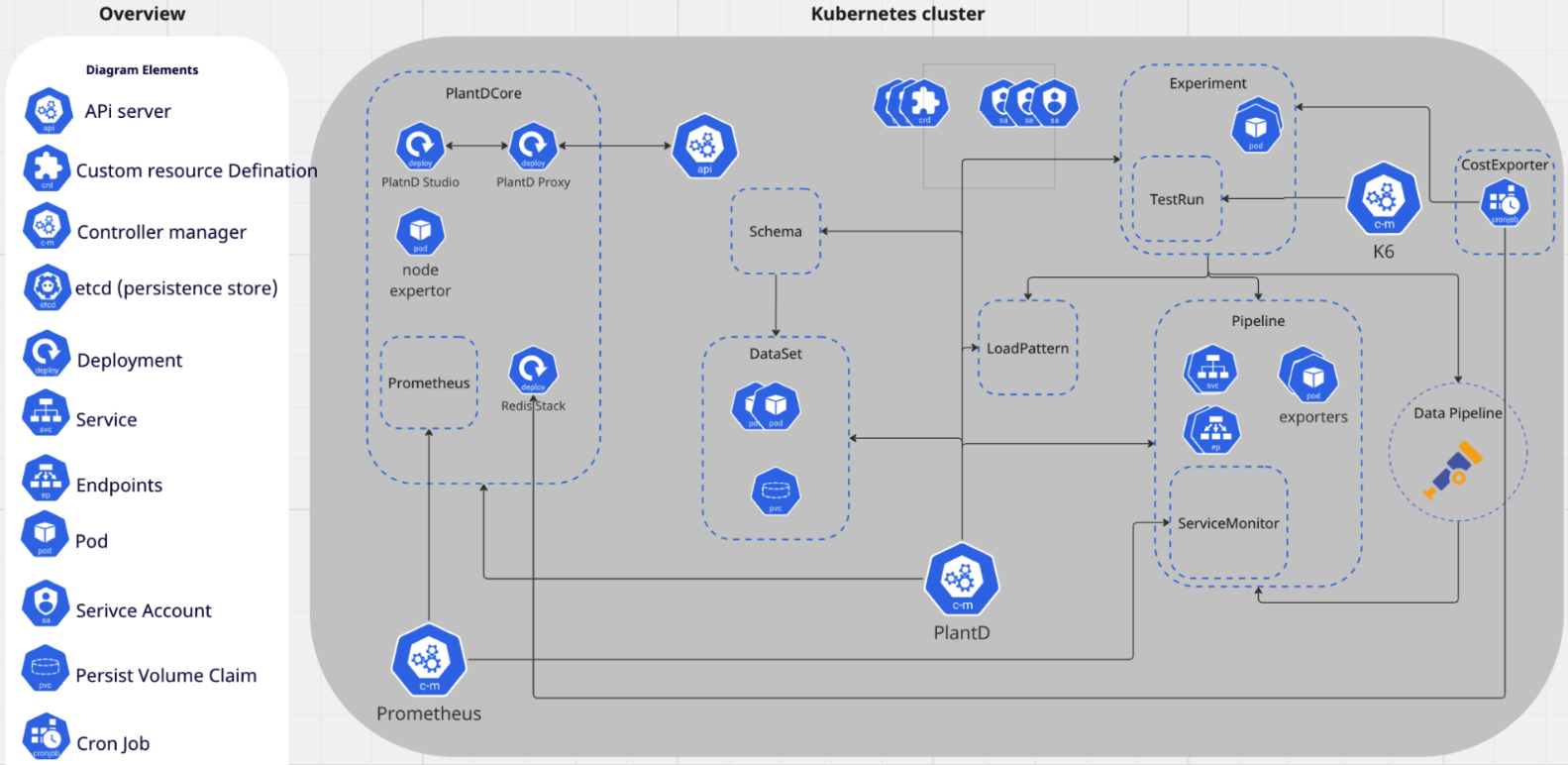}
  \caption{General system overview of \plantd with its major components, implemented as Kubernetes Custom Resources.  \plantd Core (left) manages user-facing resources such as the user interface (\plantd-Studio) and the Prometheus and Redis repositories of collected data. \emph{Schema} and \emph{DataSet} describe the data format that must be synthesized to feed a  pipeline-under-test; \emph{LoadPattern} describes the timing and quantity of data fed to the pipeline; \emph{Pipeline} describes the endpoint and protocol of a pipeline; finally \emph{Experiment} ties these together, manages a scheduled experiment, and points to the data collected from it.}
  \label{fig:architecture}
\end{figure*}

\plantd's architecture, shown in Figure~\ref{fig:architecture}, consists of a management service with API and GUI, a load generator, and a monitoring pipeline built using various open-source technologies.  

\subsection{Pipeline Under Test}
The \texttt{pipeline-under-test} is the artifact that engineers want to measure; it should consist of some endpoint to which data is sent for ingestion, one or more processing stages, and optionally some repository where data is placed after processing is complete. This last repository may be a data lake, data warehouse, or some other structure.  The pipeline may be hosted anywhere and use any technology, although the data and load generators may need to be extended for novel protocols and data formats.

\subsection{Data collection}
A \plantd user wanting ``white box'' evaluation of a pipeline may additionally add specialized logging to their code, using the open source OpenTelemetry package~\cite{OTel}.  To minimize effort for the engineers, we minimize the amount of instrumentation required: OpenTelemetry spans must be declared, logging the start time and duration of each stage of the pipeline the user wishes to measure.  \plantd provides an OpenTelemetry collector that converts these spans to metrics and provides them to an endpoint for Prometheus. 

Besides these stage entry/exit logs, infrastructure metrics of the pipeline's resources, like CPU load or memory usage, can also be collected from the cloud provider or the Kubernetes infrastructure. These can provide additional context for understanding pipeline behavior. 

\subsection{Data generator}
The data generator is responsible for mimicking realistic data the pipeline-under-test is designed to receive. It generates a quantity of data and stores it in advance of an experiment.  To generate data, \plantd uses a service built with the open source GoFakeIt\cite{GoFakeIt} package, which describes a wide variety of domain-specific data types and their ranges. This service allows the pipeline engineer to input constraints on the structure, types and value ranges of the data it generates, and the total quantity of data they expect to send. The resulting data is then stored in a DataSet object, a Kubernetes custom resource which feeds data to a Load Generator.  The Data Generator has a few common choices of data format (e.g. zipped CSV files); other formats could be easily added to the open-source codebase.

\subsection{Load generator}

\plantd uses the open-source K6~\cite{K6} tool to send pre-generated data at a controlled rate to the \texttt{pipeline-under-test}. Through the GUI, the user specifies a sequence of time spans, and the start and end data rate for each span. \plantd configures K6 to send at those rates, and linearly interpolate rates if the start and end rates differ in a segment. K6 can be scaled to handle sustained high data rates.  

\subsection{Cost data}
Measuring the cost of a pipeline-under-test can be challenging.  Although cloud providers provide detailed billing for the allocation and use of resources, pipeline costs can be difficult to distinguish from other activities in the same subscription, for two main reasons.  First, cloud subscriptions tend to encompass many small services with a proliferation of names, making it a chore to identify relevant resources.  Secondly, some resources may be shared between relevant and irrelevant tasks. For example in the case of a pipeline running in a shared Kubernetes cluster,
one node (a cloud-provided virtual machine) may be doing some tasks relevant to the pipeline-under-test, and some irrelevant tasks.  The cost of the resource must be broken down and allocated fairly. Finally, the time granularity of billing records may not be precise enough to align with the start and end of an experiment. 

For non-Kubernetes pipelines directly using cloud resources such as VMs or a mix of managed resources, the engineers measuring a pipeline must tag their pipelines-under-test to differentiate them in the cost logs from other cloud resources. \plantd's cost service then queries the cloud provider's billing information (AWS or Azure), adding cost records matching those tags to a Redis database when they become available from the provider, typically a few hours after the experiment.  

Azure and AWS provide cost logs only with the granularity of one hour, but when prorated for the length of a test, they provide us with a fairly realistic cost estimate.  For more accurate cost accounting when a pipeline is expected to have significant cost changes over time, (for example because of automated resource scaling), longer experiments can be made for which an hour is sufficient resolution.

For Kubernetes pipelines, provider billing records make the cost of an individual resource difficult to tease apart from other cluster tasks and the Kubernetes infrastructure itself. For this, \plantd uses OpenCost~\cite{OpenCost}, an open-source Cloud Native Computing Foundation project. OpenCost allocates the costs of a Kubernetes cluster to individual containers based on node resource utilization (CPU, RAM, and Persistent Disk) as well as Load Balancer attachment and other related costs. Users provision the containers of their pipeline in a distinct Kubernetes namespace, and the \plantd cost service retrieves the total cost incurred by these containers during the experiment window. We validated the cost for correctness on an AWS cluster by running experiments with several different pipelines, comparing container utilization metrics with AWS resource pricing rates as a source of truth; OpenCost was more than 95\% accurate for all experiment runs.

\subsection{Engineering Analysis}
The tools described so far allow engineers to repeatedly run identical experiments on a pipeline, so that they can observe changes in the graphs and statistics about their pipeline's behavior as the pipeline evolves.  They can also test the same iteration of the pipeline under different conditions to see how sensitive its performance and cost is to those conditions.

Graphs show the latency, throughput, and cost over time, along with a table of overall summary statistics. Latency and throughput for each instrumented stage of the pipeline are also depicted.

\subsection{Business Analysis}
The business analysis features of \plantd allow an analyst to extrapolate the measured performance of the pipeline to the load patterns predicted for the business.  It does this by creating a mathematical model (called a \emph{digital twin} in the user interface) of how latency, throughput and cost depend on load.  

Analysts create a \emph{traffic model} describing how the volume of incoming data (the ``load'') will be expected to vary hour by hour over a year. 
The 
model
extrapolates measured model behavior 
to generate a simple digital twin based on any experiment the analyst chooses.

\includegraphics[width=0.4\textwidth]{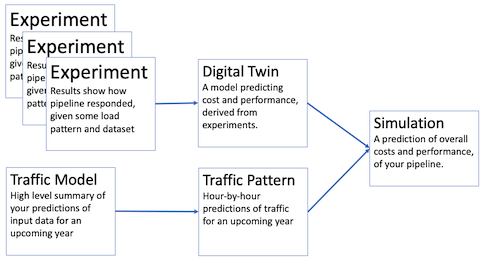}

The analyst may also specify network cost per megabyte, storage costs per gigabyte per day, and raw data storage policy, so that the simulation can also estimate network and storage costs, for storage-aware or -agnostic modes.

Using these features involves the following steps:
\begin{enumerate}
    \item \textbf{Run experiments} with various data schema mixes and load patterns to expose pipeline behaviors. 
    \item \textbf{Create model of pipeline} summarizing experimental data to predict cost, latency, and throughput based on fluctuating loads. There are predefined twin types:
    \begin{itemize}
        \item \textit{Simple Model:} Assumes fixed throughput capacity with an infinite queue.
        \item \textit{Quickscaling Model:} Assumes optimal horizontal scaling, eliminating queuing delays.
    \end{itemize}
    \item \textbf{Define traffic model} by specifying parameters (described below) to create a realistic hourly load estimate for a year.
    \item \textbf{Define network and storage costs} to refine cost predictions.
    \item \textbf{Simulate} the pipeline using the pipeline model and traffic model to estimate pipeline performance. 
    \item \textbf{Compare} simulation results of different scenarios and pipelines using the \plantd interface.
\end{enumerate}

\textit{Projected traffic load calculation}
Traffic forecasting depends on several input factors, submitted by users on a form in the GUI:
\begin{itemize}
    \item The data rate $R$ expected at the beginning of a hypothetical year. This should be the output of the analyst's own model of, for example, sales of some data-producing product, retirement/obsolescence of the product, how often it is used, and each unit's data rate.
    \item The annual growth in data rate, $G$, allowing modeling of, for example growth through sales, or decline because of obsolescence.
    \item A corrective factor $M_\mathrm{month}$ per month, capturing seasonal variation.
    \item A corrective factor $H_{\mathrm{hour},\mathrm{dow}}$ (where $\mathrm{dow}$ means day of week) for each of the 168 hours in a calendar week, capturing daily variation and weekday/weekend differences.
\end{itemize}
From these parameters, a realistic hourly load estimate $Load_h$ is projected for each hour $h$ in a year:

\[
R \times (1+\frac{\mathrm{day of year}(h) \times G}{365}) \times H_{\mathrm{hour}(h),\mathrm{dow}(h)} \times M_{\mathrm{month(h)}}
\]

The business analysis features facilitate comprehensive cost and performance forecasting, enabling both engineers and business analysts to make informed decisions about pipeline scalability and efficiency.

The analyst may also specify service-level objectives, taking the form of a measurement type (currently either latency or error rate), a maximum limit (seconds for latency, or percent per hour for error rate), and a proportion of hour violations.  An example SLO would be that processing latency may not exceed 100 seconds more than 0.1\% of the time.

Using the set of experiments (which should be on the same pipeline), the system generates a model of the pipeline's performance under different conditions. We currently generate a simple proof-of-concept implementation of this model: using a single experiment, the model assumes a fixed cost of resources, uses the total time to fully process all the records in the generated load, and calculates the apparent sustained throughput; it then assumes FIFO queuing whenever input load exceeds capacity. No synthetic data is actually processed; only the load shape is used, so the simulation is quite fast. Our design allows contributors to add more sophisticated models; the model-training module is provided with a full history of data from all the experiments, so that more detailed heuristics or machine-learned models could easily be introduced.

The GUI allows browsing of high level statistics about cost and performance, as well as detailed results from a simulation. Comparing simulations side by side allows business analysts and engineers to compare architectural options under different data load projections.

\begin{figure}
\includegraphics[width=0.4\textwidth]{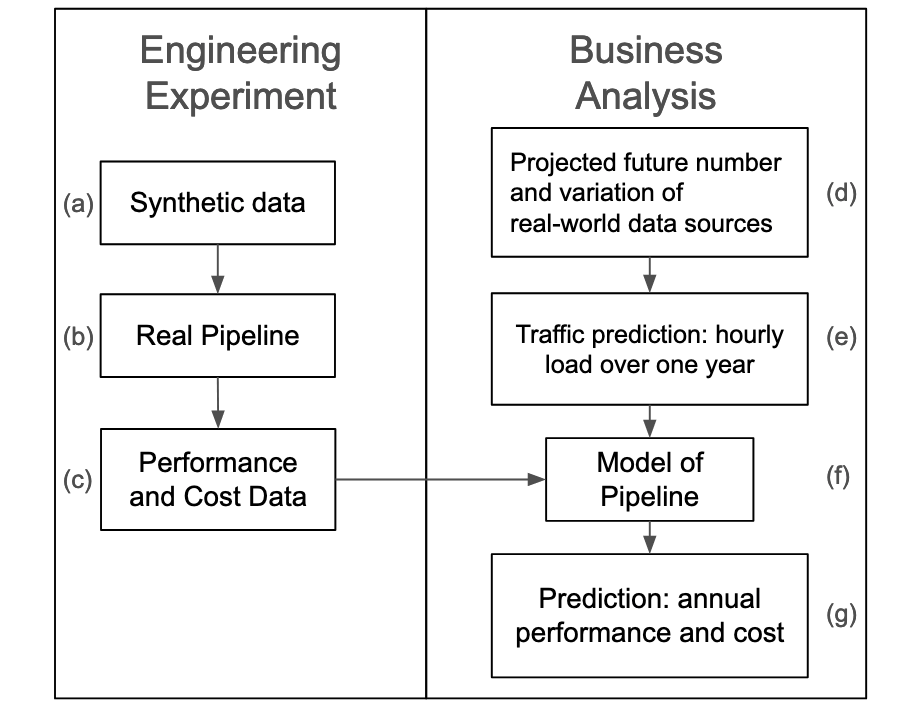}
  \caption{Conceptual flow of the research method. Engineering experiments (left) send synthetic data to a real pipeline, and measure its characteristics.  Business analysis (right) models pipeline cost and performance, and applies business projections to extrapolate performance over a future year.}
  \label{fig:method}
\end{figure}

\section{Methods}
\label{sec:method}
Figure~\ref{fig:method} summarizes the procedure used to produce the results.  We implemented and deployed cloud three engineering iterations of a data pipeline-under-test to AWS (Fig.~\ref{fig:method}b).
For each of these we describe the performance and cost data (Fig.~\ref{fig:method}c, Table~\ref{tab:experiment-summary}), which are intended to be immediately useful as benchmarks of pipeline behavior as engineers develop a pipeline and compare different architectures, configurations, and versions of their code.  

Next we demonstrate the business analysis facilities of \plantd, by answering business-related what-if questions:
\begin{itemize}
    \item What would the cost and performance impact be if increased car sales put 50\% more cars on the road by the end of the year?
    \item What would be the cost impact of doubling our data retention policy from 3 to 6 months?
\end{itemize}

In the remainder of the methods section, we describe the pipelines-under-test we will use for the engineering experiments, along with their motivating context, and the traffic models we will use for the business analysis.

\subsection{Example pipelines-under-test}
We began with a simplified version of a data pipeline to ingest streaming sensor information from a test fleet of Honda's connected vehicles. This fleet was used to demonstrate various V2X (Vehicle-to-Everything) capabilities in Ohio.

The pipeline processes data collected from Honda’s test vehicle fleet and translates it into data that could help study various in-vehicle and out of vehicle characteristics to create a safe mobility ecosystem. Currently, Honda has a small fleet of vehicles that collect a suite of metrics on a vehicle that characterize how it is being driven and operated. We generated raw test data modeled after the real data Honda receives from these vehicles daily.

The data takes the form of a stream of zip files.  Each represented a data transmission from a single car, and contains five files in a custom binary format representing data from five different automotive subsystems, such as engine status, location, and speed.  

The example test pipeline has three stages, based on a real pipeline within \honda: an unzipper, a parser, and a database-insertion phase.  Functionality is stripped down from the real-world pipeline to capture the resource-intensive parts of the processing:

First, \textbf{\texttt{unzipper\_phase}} receives zip files from an HTTP endpoint which are uploaded to AWS S3 for persistence. This stage then unzips the zip files, and adds the binary files to a Kafka queue for the next stage of processing.

Next, \textbf{\texttt{v2x\_phase}} reads the binary files from the Kafka queue and converts the custom binary telematics files to parquet format, a standard data format for pipeline processing; it then backs up the data to AWS S3 and also adds the parquet data to a Kafka queue for the final stage of processing.

Finally, \textbf{\texttt{etl\_phase}}, triggered by a Kafka message, reads each parquet file, processes the raw data records and adds the processed records, scrubbed of missing or bad data, into a MySQL RDS database on AWS.

We set up and tested the pipeline (called \modelBw in the sections below), then made exploratory changes based on \plantd's measurements (\modelNbw and \modelCpu). We describe these changes briefly in the narrative of the results section (Sec.~\ref{sec:results}), but the technical details are not important; they are included to illustrate that \plantd's usefulness for pinpointing the effects of engineering problems and demonstrating the impact of fixes.

\subsection{Traffic projections for business analysis}
As a context for predicting cost and performance of operating the three variant pipelines under realistic business projections, we describe two projections: the first, \trafNom projection, assumes a stable population of cars on the road over a future year, and a second, \trafHigh projection, starts at the same population, but assumes 50\% growth in installed vehicles. These forecasts were based on the following back-of-envelope assumptions:
\begin{itemize}
    \item In the \trafNom condition, we assume an arbitrary number of 250,000 instrumented cars on the road, 50\% of whose drivers opt into telematics, and 4\% being on the road at any given time (assuming each car is driven about 1 hour a day), and sending one file per hour of telematics: this gives us about 5000 records being received per hour on average, at the start of the year. We project no net growth over the target year. 
    \item In the \trafHigh condition, we assume 50\% growth.
    \item We multiply each record-per-hour value by an adjustment for the month (ranging from 0.84 in January to 1.14 in August) to account for seasonal variation, and for each hour of the week (ranging from 2.26 for Friday at 8pm, to 0.04 for Weds at 6am).  These were abstracted from measurements from a Honda test program.   
\end{itemize}

\subsection{Modeling pipeline behavior}
Turning to business analysis, we used \plantd to build mathematically models of the three pipeline variants, in order to apply them to traffic projections.  

These models predict cost, latency and throughput of each pipeline over time as a function of an input stream of varying data input rates. 
Note that data is available for more sophisticated models to be built: CPU and memory usage, for example, are available to be used to separate fixed costs from variable cost based on resource assumptions; the dynamic behavior of the pipeline in response to load changes could be modeled more closely with machine learning; or autoscaling behavior could be predicted by wrapping a fixed model based on measurements with a autoscaling rules. 

However, while \plantd's open-source architecture allows for arbitrary modeling techniques to be added, for this proof-of-concept we use a very simplistic model, opting for explainable parameters and a rough approximation of pipeline behavior over the two traffic projections.

The parameters of the models are shown in Table~\ref{tab:twinmodel-summary}:
\begin{itemize}
    \item \texttt{max rec/s}: The pipeline ingests up to this many records each second, drawing first from any queued data not previously processed, then from the incoming data.
    \item \texttt{\$/hr}: This model assumes resource costs are fixed per hour, and do not change based on load.
    \item \texttt{avg. latency}: Latency of a record to be completely processed through the entire pipeline assuming no queuing.
    \item \texttt{policy}: The queue is assumed to be processed in FIFO order.
\end{itemize}

The models' parameters reflect the experimental finding that the \modelNbw pipeline has significantly higher throughput, but at a higher cost per hour than the other two.  In fact, dividing those two parameters shows that the \modelNbw pipeline is modeled as being about three times as expensive per record processed (\$0.00032) as the \modelNbw pipeline (\$0.00012).  The \modelCpu model is similar at (\$0.00011).

\begin{table}[]
\begin{tabular}{llllll}
Model & max rec/s & \$/hr & avg latency & policy\\ \hline
blocking-write & 1.95 & 0.82 & 0.15 & fifo\\
no-blocking-write & 6.15 & 7.03 & 0.06 & fifo\\
cpu-limited & 0.66 & 0.27 & 0.29 & fifo\\
\hline

\end{tabular}
\caption{Parameters of three twin models, derived from the three experiments.}
\label{tab:twinmodel-summary}
\end{table}


\subsection{Forecasting the demand}

\begin{figure}
\includegraphics[width=0.5\textwidth]{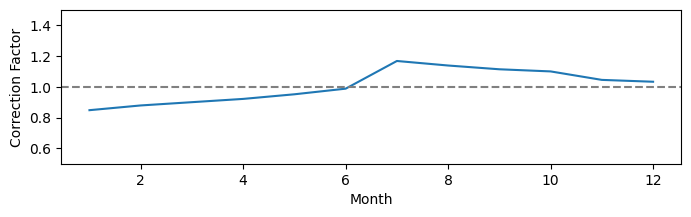}
\includegraphics[width=0.5\textwidth]{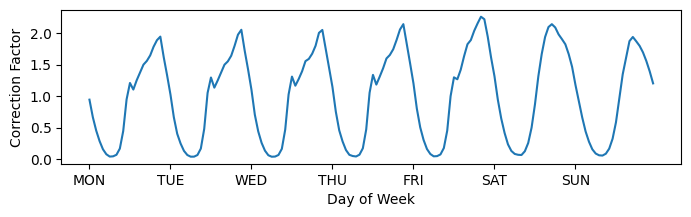}
\includegraphics[width=0.5\textwidth]{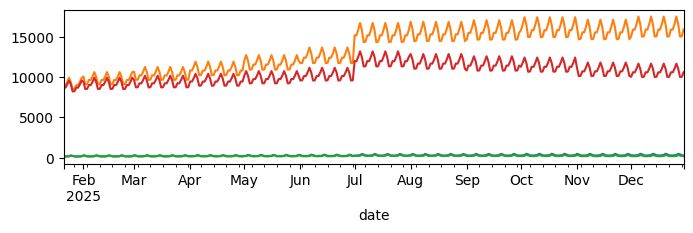}
  \caption{Correction factors for each month (top), for each hour within a week (center), and resulting \trafNom and \trafHigh projections (bottom), assuming no net change (\trafNom) or growth (\trafHigh). Red is daily maximum of \trafNom project; orange is daily maximum of \trafHigh projection, and green is the daily minimum of both projections.}
  \label{fig:projections-minmax}
\end{figure}

Next we ran simulations under the \trafNom and \trafHigh projections described in Section~\ref{sec:method}.  Figure~\ref{fig:projections-minmax} shows the traffic predictions.

The projections are based on an assumption of beginning of year traffic in records per second (we used 3.5 rps for both projections), an annual growth factor (1.0 for the \trafNom case, meaning no growth, and 1.5 for the \trafHigh case, meaning a net 50\% increase in load), and corrections factors for each month (Figure~\ref{fig:projections-minmax}(top)), and for each hour within a calendar week Figure~\ref{fig:projections-minmax}(top)).

These assumptions and correction factors were back-calculated from actual measured \honda data from a very narrow sample of cars; we assume that business analysts will upload more realistic, application-specific predictions into the \plantd-studio interface.

\begin{table*}[]
\begin{tabular}{lllllllll}
run & cost (\$) & \multicolumn{3}{c}{latency (s)} & \multicolumn{2}{c}{thruput (rec/h)} & \% latency met & SLO met\\ 
 &  & median & mean & backlog & mean & max &  & \\ 
\hline
nom block & 71.87 & 258.59 & 2938.41 & 6690.64 & 5035.8 & 7024.39 & 97.02 & True\\
nom non-block & 614.19 & 0.06 & 0.06 & 0.0 & 5037.29 & 13191.79 & 100.0 & True\\
nom cpu-lim & 50.56 & 14528142.34 & 16032353.34 & 35130437.72 & 2378.86 & 2380.17 & 0.17 & False\\
high block & 74.71 & 20400.13 & 280566.55 & 1247902.13 & 6096.8 & 7024.39 & 47.84 & False\\
high non-block & 614.19 & 0.06 & 0.06 & 0.0 & 6375.49 & 17502.77 & 100.0 & True\\
high cpu-lim & 63.98 & 18521842.1 & 21802654.3 & 52813607.51 & 2378.92 & 2380.17 & 0.17 & False\\
\hline
\end{tabular}
\caption{Summary of simulations of various pipeline models and traffic forecast pairs.  Latency backlog is the amount of time needed to process backlog of unprocessed records at the end of the year.}
\label{tab:simulation-summary}
\end{table*}

\begin{figure}
\includegraphics[width=0.5\textwidth]{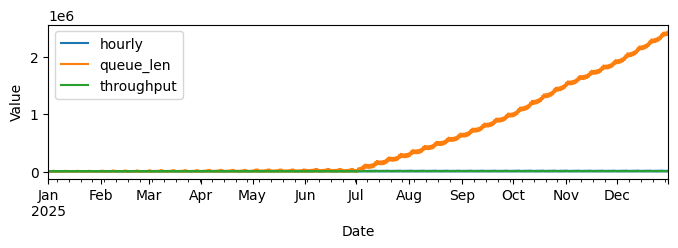}
  \caption{Whole-year simulation of \modelCpu model under the \trafNom projection.  It predicts that queue length (and hence latency) would grow out of control starting in July when traffic is predicted to pick up, and the pipeline would never catch back up.}
  \label{fig:runaway}
\end{figure}

\begin{figure}
\includegraphics[width=0.5\textwidth]{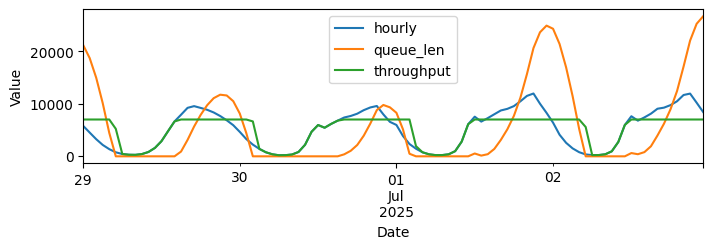}
  \caption{Excerpt of simulation of \modelBw model under the \trafNom projection.  Incoming load (blue) exactly matches pipeline throughput early each day (green) until throughput maxes out at around 7000 rec/hr. At that point the queue (orange) begins to grow.  When load falls back to a manageable rate (where blue and green lines cross), the pipeline makes headway on the backlog, and the queue begins to descend from its peak for the day.}
  \label{fig:projections}
\end{figure}

For the business analysis experiments, we further start with the assumption that network transmission from the car costs .02\textcent/MB, storage costs are 1\textcent/GB/day, and raw data is retained for 3 months after it is collected.

\begin{table*}[]
\begin{tabular}{llllllll}
experiment & mean throughput & mean latency & median latency & exp. length (s) & total cost (\$) & cost/hr (\$)\\ \hline
blocking-write & 1.95 & 0.15 & 0.15 & 1230.0 & 0.28 & 0.82\\
no-blocking-write & 6.15 & 0.06 & 0.06 & 390.0 & 0.76 & 7.03\\
cpu-limited & 0.66 & 0.29 & 0.29 & 3630.0 & 0.28 & 0.27\\
\hline
\end{tabular}
\caption{Experiment results for three different pipeline variants.  Cost and cost per hour are in cents.}
\label{tab:experiment-summary}
\end{table*}

\section{Results}
\label{sec:results}

\begin{figure*}[htbp]
    \centering
    \begin{subfigure}[b]{0.32\textwidth}
        \includegraphics[width=\textwidth]{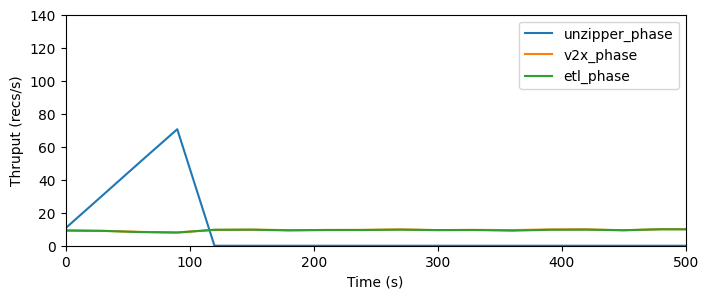}
        \caption{\modelBw  Throughput}
        \label{fig:bad-throughput}
    \end{subfigure}
    \hfill
    \begin{subfigure}[b]{0.32\textwidth}
        \includegraphics[width=\textwidth]{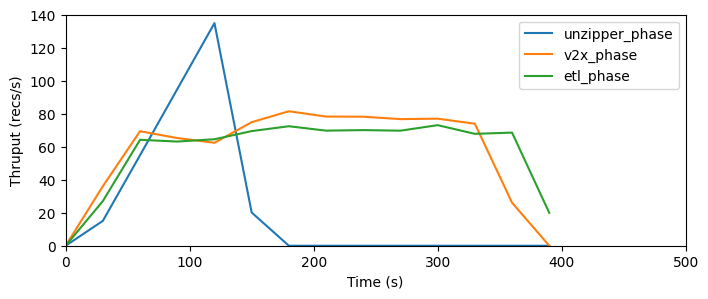}
        \caption{\modelNbw  Throughput}
        \label{fig:good-throughput}
    \end{subfigure}
    \hfill
    \begin{subfigure}[b]{0.32\textwidth}
        \includegraphics[width=\textwidth]{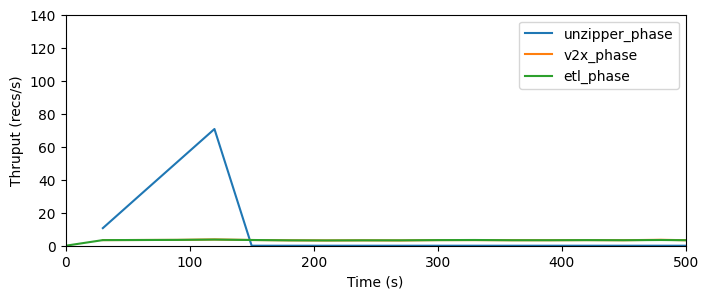}
        \caption{\modelCpu  Throughput}
        \label{fig:fixed-throughput}
    \end{subfigure}

    \begin{subfigure}[b]{0.32\textwidth}
        \includegraphics[width=\textwidth]{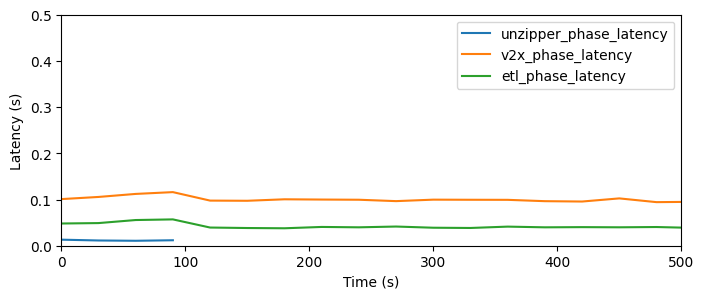}
        \caption{\modelBw  Latency}
        \label{fig:bad-latency}
    \end{subfigure}
    \hfill
    \begin{subfigure}[b]{0.32\textwidth}
        \includegraphics[width=\textwidth]{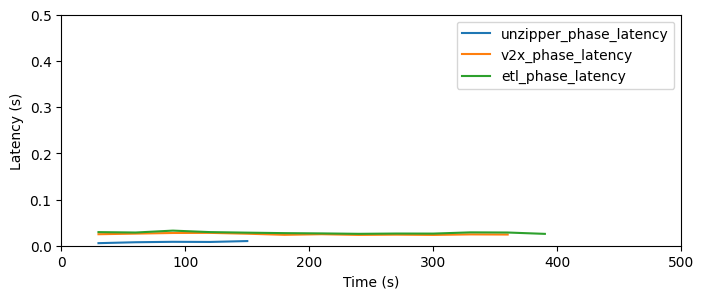}
        \caption{\modelNbw  Latency}
        \label{fig:good-latency}
    \end{subfigure}
    \hfill
    \begin{subfigure}[b]{0.32\textwidth}
        \includegraphics[width=\textwidth]{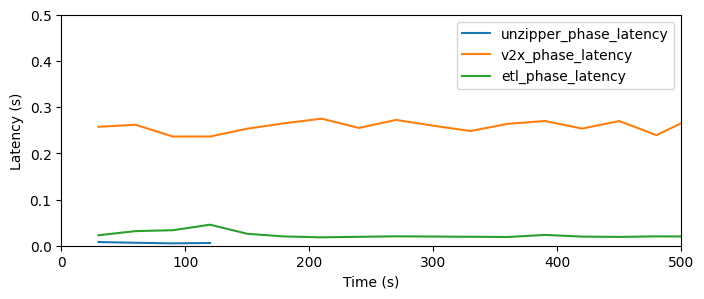}
        \caption{\modelCpu  Latency}
        \label{fig:fixed-latency}
    \end{subfigure}

    \caption{Throughput and latency of models of three pipeline variants: \modelBw, \modelNbw, and \modelCpu.  Graphs break down metrics by the pipeline's three phases.  The blue spike represents rapid processing by the first phase, but differences among the pipelines were most evident in their second phase, \texttt{v2x\_phase}. Graphs are cut off at 500s; see Table~\ref{tab:experiment-summary} for the actual experiment lengths.}
    \label{fig:models-comparison}
\end{figure*}

\subsection{What-if: Comparing engineering variants}
We ran our first experiment with the \modelBw variant of the telematics pipeline-under-test, using a load pattern lasting 120 seconds, ramping up linearly from 0 to 40 records per second, in order to reveal the point when the pipeline could no longer keep up with load.  The left column of Figure~\ref{fig:models-comparison} shows the latency (top) and throughput (bottom) for the each of the three stages of this pipeline; the colored lines represent the three stages of each pipeline.  The top row of Table~\ref{tab:experiment-summary} summarizes the experiment's results.

The throughput graph (top left) shows that the first \texttt{unzippper\_phase} has the highest throughput, keeping up with the load as quickly as it is generated.  We can hypothesize that the second stage (\texttt{v2x\_phase}) is the system's bottleneck, since the third stage (\texttt{etl\_phase}) is able to keep up with it, hence the throughput graph is overlaid with it and mostly indistinguishable in the graph.

The latency graph (bottom left of Figure~\ref{fig:models-comparison}) bears this out, showing that the first phase has negligible latency, and completes almost immediately, while the second stage takes the longest.

After observing this relationship, we examined the pipeline's code to look for a way to mitigate this first observed bottleneck.  The \texttt{v2x\_phase} was writing each file received to Amazon S3 blob storage, and we realized this was a blocking write of duplicate information: we suspected that this was the main cause of the high latency of the operation.  We tested this by modifying the code to remove the write (variant \modelNbw), set up the new variant as a separate pipeline in the cloud, and ran an identical experiment on it using \plantd.

The middle column of Figure~\ref{fig:models-comparison} shows the results; \texttt{v2x\_phase} is now of comparable latency to \texttt{etl\_phase}, and the throughput is considerably higher.

Note that for white-box analysis of data \emph{within} particular pipeline stages, we focus on differences between pipeline variants rather than absolute numbers.  For an arbitrary pipeline, the number of records processed by one phase may not match the number processed by the next, since the processing may for example involve splitting or joining records, or outputting records that somehow crosscut the input records.  As a general-purpose tool, \plantd can make no assumptions about the relationship between the number of records sent by the load generator, and records processed by each phase. In our case, the first \texttt{unzipper\_phase} extracts five automotive subsystem data files from each ingested zip file from the vehicle, and passes each to the second phase.  So for example the \modelBw throughput (upper left of Fig.~\ref{fig:models-comparison}) appears to be about 10 records per second, five times what the table shows as a mean throughput of 1.95 records per second.

In a second version of the telematics pipeline (\emph{no-blocking-write}), we removed the blocking write to cloud blob storage; this resulted in apparently higher performance.  This second version handles the same load in less than ten minutes.

In a third version (\modelCpu), we set Kubernetes to deliberately throttle the CPU of the second stage of the \modelNbw model, to verify that it would have a similar effect as the blocking write did. 

These graphs and figures, along with other measured observables like CPU and network load, may be sufficient for engineers working on data pipelines.

\subsection{What-if: Increased car sales}
Next, we applied the three models to both traffic forecasts, setting a service level objective (SLO) of processing all records within 4 hours, 95\% of the time. Table~\ref{tab:simulation-summary} shows the costs and performance of all six combinations: the SLO was met in three of the six cases. Although it might not occur to an engineering team to run such a business simulation on a model like \modelBw that was later optimized, the results of simulations show that it may be a better choice under some assumptions. 

The  \modelCpu model was the cheapest per hour, but as demand rose midway through the year in both forecasts, the model could not keep up; the simulation shows it getting further and further behind after July, until by the end of the year it has a backlog of about 406 days of work, in the \trafNom projection (Figure~\ref{fig:runaway}), and 611 in the \trafHigh projection. (We assign a cost to these end-of-year backlogs by multiplying the queue length at the end of the year by the cost of the pipeline per hour, capturing the cost of, for example, spinning up duplicate pipelines to process the backlog).

The \modelBw and \modelNbw pipelines both handle the \trafNom load within the SLO; \modelBw does so just barely (latency violates the rule 3\% of the time) but much more economically than the \modelNbw.

If demand grows in the way modeled by the \trafHigh projection, the \modelBw model can no longer keep up with the load; the latency SLO will be continuously violated after the middle of the year.  However notice that even after processing the huge backlog of records not processed at the end of the year, the \modelBw model is significantly cheaper; suggesting that adding some autoscaling to this model might be a better choice.

Figure~\ref{fig:projections} illustrates how the simulation works, depicting \texttt{blocking-write}'s performance on the \trafNom traffic projection. During a high-traffic period, the pipeline could not quite keep up with the peak incoming load in real time, but it could recover at night.  The green line shows a queue buildup, which is later handled when the load dips back below the pipeline's capacity.

\begin{table}[]
\centering
\begin{tabular}{lcccc|cc}
\hline
 & \multicolumn{4}{c|}{3 mo. retention} & \multicolumn{2}{c}{6 mo. retention} \\ 
month & cloud & net & storage & total & storage & total \\
\hline
1 & 52.3 & 0.43 & 7.78 & 60.51 & 7.78 & 60.51\\
2 & 47.24 & 0.4 & 20.7 & 68.35 & 20.7 & 68.35\\
3 & 52.3 & 0.46 & 38.55 & 91.3 & 38.55 & 91.3\\
4 & 50.61 & 0.45 & 46.0 & 97.07 & 53.29 & 104.35\\
5 & 52.3 & 0.49 & 48.88 & 101.66 & 72.06 & 124.85\\
6 & 50.61 & 0.48 & 48.68 & 99.78 & 86.75 & 137.85\\
7 & 52.3 & 0.59 & 53.33 & 106.23 & 100.93 & 153.83\\
8 & 52.3 & 0.58 & 57.3 & 110.18 & 106.26 & 159.15\\
9 & 50.61 & 0.55 & 58.17 & 109.33 & 106.96 & 158.12\\
10 & 52.3 & 0.56 & 60.41 & 113.27 & 114.03 & 166.89\\
11 & 50.61 & 0.51 & 57.09 & 108.22 & 112.7 & 163.83\\
12 & 50.68 & 0.51 & 55.67 & 106.86 & 113.99 & 165.18\\
\hline
total & 614.19 & 6.01 & 552.56 & 1172.76 & 934.0 & 1554.2\\
\end{tabular}
\caption{Monthly costs for the nominal no-blocking model.}
\label{tab:monthly-summary}
\end{table}

\subsection{What-if: Changing a data storage policy}
To investigate the cost implications of data storage policy, we
reevaluated the cost of doubling the data retention period from 3 to 6 months. Table~\ref{tab:monthly-summary} shows the monthly cost of the nominal/blocking model along with network and storage costs.  \plantd 
calculates the storage costs by simulating the accumulation and aging of data. Using a rolling retention window, data builds up in storage daily and is automatically removed once it surpasses the retention period. This approach allows us to calculate the total amount of data stored on each day, which we then use to estimate storage costs by multiplying the daily storage volume by the storage cost. 

Of interest are both the yearly cost, and the steady-state average cost after the retention period expires for the first day (since the initial retention period holds data as the pipeline ramps up into operation, and is thus unrepresentative of ongoing costs).

\section{Discussion}
\label{sec:discussion}
Our first iteration of the example pipeline-under-test had what the team considered a poor design choice, adding a blocking cloud operation in the middle of a data pipeline phase, despite our team's familiarity with commercial cloud technologies. Even for experts, cloud computing tools such as load generators, Kafka, EMR, virtual machines, and database engines are not only complex to configure in themselves, but their configurations impact each others' performance, and their efficiency is dependent on the characteristics of data fed into them.  When a configuration finally ``works'' and begins producing correct results at some rate, it is not immediately apparent how efficiently the pipeline is using its resources, whether it needs tuning, and if so how.  

Fixing this blocking call seemed obvious to us, since it increased the throughput of the pipeline, and rerunning the experiment with \plantd gave us immediate verification of that improvement.  Yet the economic modeling shows a more nuanced picture: something about the way the \modelNbw model operates is quite cost-inefficient for the increase in speed, to the point where simply duplicating the \modelBw pipeline would boost throughput more economically.

Outside of extremely routine deployments, we don't believe cloud engineering teams can expect to get things right the first time, without a cycle of measurement and refinement.
Our goal in this research is to show how a tool like \plantd can enable exploration of the cost and performance trade-offs of different design choices and implementation variants, by showing what it reveals to a team about the options facing them.

\subsection{Product Engineering and Business Insights}
Through the development process, \plantd produces end-to-end metrics of latency, throughput, and cost, and provides evidence of the scalability limits of pipelines under test.  This information is most easily gathered using IT-centric metrics, like bytes or records per second, but can just as readily be cast in business-centric terms by converting data rates to implied numbers of cars and the size and frequency of data packages they transmit.

This early, easy availability of business-relevant metrics can prevent unpleasant surprises, by providing business development experts and managers visibility from the start and well into the development progress towards a pipeline, helping them predict whether the pipeline will eventually cost more to operate than the revenue it is supposed to generate.

Additionally, if a wind tunnel produces eye-catching, understandable visualizations that encompass a standard suite of performance and cost measures, easily convertible to units relevant to the business (e.g. number of data-generating mobility devices such as cars, rather than data ingestion traffic bits per second, for \honda) then these visualizations will be useful within the organization. 
Circulating this suite of measurements could help educate management and business planners towards a better understanding of how performance and cost trade off in pipelines, and how these trade-offs interact with business goals, on par with other aspects of a business. The research team at Honda has observed that different stakeholders such as IT, business analysts and product leaders interpret results differently. The different roles lack common terminology and shared frameworks needed to consider such trade-offs alongside other more traditional considerations.

Another advantage of a high-level business-facing suite of measurements is that they can serve as a common set of \emph{boundary objects}~\cite{Star1989}; that is, a shared set of concepts and artifacts that enable communication between collaborators with different specialties, such as business decision makers and engineers.  Common language helps engineers understand what management's priorities and sensitivities are, and thus make more confident architectural decisions.  

As engineering teams evolve and learn about new cloud based technologies, cost very quickly becomes complex to calculate.
Each cloud provider has its own mechanisms for cost calculations, and each provided service affords different technologies, solutions, and optimizations. To untangle each service's contributions to the cost is a very complex process. A wind tunnel approach allows engineering teams to measure and report actual costs, and frame them in a way that business developers can digest. This in turn gives engineering more freedom to explore innovative options with different cost models.

If many engineering teams in an organization used \plantd, rather than each having to develop their own idiosyncratic metrics, it would allow comparable, mutually intelligible cost and performance data to be shared throughout the organization, described in a common metrics language. 

\section{Future Work}
\label{sec:future_work}
\plantd is designed to allow more sophisticated traffic and pipeline modeling, and to relatively easily add new graphs and tables of results to the interface.  Interesting improvements in the modeling might include statistically characterizing burstiness of real-world traffic, to model very short-term peaks, or using machine learning for more accurate pipeline models.  We also expect that data pipeline paradigms may shift towards cloud-edge architectures, distributing intelligence across computing layers. This shift suggests adding ways of specifically comparing tradeoffs in distribution of ML pipelines depending on bandwidth, reliability, costs, and power constraints of devices in complex distributed architectures.

\section{Conclusions}
\label{sec:conclusions}
We have demonstrated the use of \plantd data pipeline wind tunnel in the domain of automotive telematics, but we believe it will be equally useful more generally in applications such as IoT, movie streaming, and beyond. A Data Pipeline Wind Tunnel is an efficient way for engineers building any sort of data pipeline to gather a coherent, complete, legible set of cost and performance requirements, all at once, every time they test their pipeline. Making these available within an organization during development, as \emph{leading indicators} avoids the risk that comes from only measuring them in production as \emph{lagging indicators}.
Wind tunnel metrics make the progress of engineers visible and understandable by management,
make comparison among different architectures and technology easier by standardizing measurements, and allow an organization to make informed choices about the flood of data coming their way: what to keep, what to process, and what to throw away.

\section*{Acknowledgment}
\label{sec:acknowledgment}
This work was a collaboration with \honda, and supported by a grant from \honda. We also thank the many CMU students and staff and Honda staff who contributed to the project, including Murphy Austin, Akshit Bhalla, Kanishka Bandaru, Kent Broestl, Yizhou Cheng, Brian Coy, Ben Davis, Frank He, Shicheng Huang, Jeahong Hwang
Siddharth Kandimalla, Abhishek Parikh, Jay Patel, Anup Kumar Prakashan, Divya Prem, Arjun Manjunatha Rao, Sarthak Tandon, Abby Vorhaus, Yuchen Wang, Xiaoyu Zhang, Tom Zhu, and Ziteng Shu.


\end{document}